\documentclass{article}

\usepackage{microtype}
\usepackage{graphicx}
\usepackage{subfigure}
\usepackage{booktabs} 

\usepackage{hyperref}

\usepackage{enumitem}

\usepackage[accepted]{icml2025}

\usepackage{amsmath}
\usepackage{amssymb}
\usepackage{mathtools}
\usepackage{amsthm}

\usepackage[capitalize,noabbrev]{cleveref}

\theoremstyle{plain}
\newtheorem{theorem}{Theorem}[section]
\newtheorem{proposition}[theorem]{Proposition}

\theoremstyle{definition}

\newtheorem{assumption}[theorem]{Assumption}
\theoremstyle{remark}

\usepackage[textsize=tiny]{todonotes}

\icmltitlerunning{MARINE: Theoretical Optimization and Design for Multi-Agent Recursive IN-context Enhancement}

\begin{document}

\twocolumn[
\icmltitle{MARINE: Theoretical Optimization and Design for \\
Multi-Agent Recursive IN-context Enhancement}



\icmlsetsymbol{equal}{*}
\icmlauthor{Hongwei Zhang}{comp}
\icmlauthor{Ji Lu}{comp}
\icmlauthor{Yongsheng Du}{comp}
\icmlauthor{Yanqin Gao}{comp}
\icmlauthor{Lingjun Huang}{comp}
\icmlauthor{Baoli Wang}{comp}
\icmlauthor{Fang Tan}{comp}
\icmlauthor{Peng Zou}{comp}
\begin{icmlauthorlist}

\end{icmlauthorlist}

\icmlaffiliation{comp}{Zhongxing Telecom Equipment (ZTE), China}

\icmlcorrespondingauthor{Ji Lu}{AIM@zte.com.cn}

\icmlkeywords{Multi-agent systems, large language models, online learning, in-context learning}

\vskip 0.3in
]



\printAffiliationsAndNotice{}  

\begin{abstract}

Large Language Model (LLM)-based agents demonstrate advanced reasoning capabilities, yet practical constraints frequently limit outputs to single responses, leaving significant performance potential unrealized. This paper introduces MARINE (Multi-Agent Recursive IN-context Enhancement), a theoretically grounded framework that reconceptualizes test-time reasoning as iterative refinement of a persistent reference trajectory, fundamentally departing from conventional one-shot or multi-sample paradigms. The MARINE refinement operator systematically converts a base model's pass@N capabilities into near-optimal pass@1 performance. Rigorous theoretical analysis establishes that minimal feasible batches maximize expected performance gains under fixed invocation budgets, while logarithmically growing batch schedules ensure continuous improvement without computational constraints. Comprehensive evaluation on the BrowserComp-ZH benchmark demonstrates state-of-the-art results, with a 685B-parameter implementation achieving 46.0\% pass@1 accuracy. Meanwhile, MARINE establishes a new paradigm for parameter-efficient reasoning: an 80B-parameter model augmented with MARINE matches the performance of standalone 1000B-parameter agents, reducing parameter requirements by over an order of magnitude. Notably, within a fixed computational budget, the proposed MARINE delivers higher-quality samples to alignment and optimization processes than traditional sampling-and-ranking strategies. Consequently, it has great potential to boost post-training efficiency.
\end{abstract}

\section{Introduction} 

\subsection{Background and Motivation}

Large Language Model (LLM)-based agents achieve strong performance on complex reasoning and searching tasks \cite{li2025system,xu2025towards,chen2025towards}. In practice, many systems are restricted to a single interaction round. Under this constraint, the pass@1 performance (probability of success with a single sample) can be substantially lower than the pass@N performance (probability of success with $N$ samples) \cite{brown2024large,zhang2025and}. \par

Existing methods for enhancing reasoning performance can be grouped into three paradigms. The first comprises multi-sampling-and-selection schemes, such as Self-Consistency (SC) \cite{wang2022self} and Best-of-N (BoN) \cite{gui2024bonbon}. These methods generate multiple reasoning chains and select the most consistent answer across samples, thereby mitigating the brittleness of any single chain that may drift away from the correct solution. Although such techniques often yield substantial gains, their effectiveness is highly sensitive to the sampling budget and lacks a theoretical guarantee that performance is strictly monotone in $N$. Particularly, when all sampled trajectories are of only mediocre quality, the pass@1 accuracy can be nearly indistinguishable from standard single-sample decoding.  \par

The second paradigm comprises self-correction and reflection-based reasoning methods. Representative approaches such as Self-Refine implement test-time iterative loops of ``generate–self-evaluate–self-revise” to iteratively improve model outputs \cite{madaan2023self}. Tree-of-Thoughts (ToT) further formulates reasoning as an explicit tree search over intermediate ``thoughts”, enabling branching and backtracking over partial solution sketches to induce more structured problem-solving trajectories \cite{yao2023tree}. Despite their success, these methods remain largely empirical. First, they lack a precise formalization and quantitative guarantees that each revision step actually improves the quality of the trajectory. Second, prior work has shown that single-agent reflection is vulnerable to Degeneration-of-Thought (DoT): once a model develops overconfidence in a candidate solution, subsequent reflection rarely escapes the established reasoning path \cite{liang2024encouraging}. \par

The third paradigm focuses on parameter optimization through Reinforcement Learning (RL). RL from Human Feedback (RLHF) methods \cite{bai2022training}, together with algorithms in the style of Group Relative Policy Optimization (GRPO) \cite{shao2024deepseekmath}, update model parameters using human preference signals or automatically constructed reward functions. These approaches have achieved substantial gains on various reasoning benchmarks \cite{shao2024deepseekmath}. However, their transferability to novel tasks remains limited, as policies often require realignment when deployed in new contexts, restricting their general applicability.  \par

In summary, existing methods primarily optimize model parameters rather than enhancing in-context reasoning processes, or rely on multi-sample decoding and self-correction strategies devoid of quantitative theoretical characterization regarding reasoning evolution. These fundamental limitations motivate our central research question: \textbf{without parameter updates, can a structured multi-agent refinement framework systematically transform an LLM's pass@N capability into reliable pass@1 performance while providing theoretical guarantees of monotonic improvement throughout the refinement process?}

\subsection{Contributions}

MARINE (Multi-Agent Recursive IN-context Enhancement) treats test-time reasoning as iterative refinement around a persistent reference trajectory instead of one-shot decoding or independent multi-sample selection. The framework operates purely at inference time and is model-agnostic. The main contributions are fourfold.

\begin{itemize}[itemsep=0pt, topsep=0pt, parsep=0pt]
    \item \textbf{Trajectory Refinement Paradigm}. The framework reconceptualizes test-time reasoning as an iterative optimization over reference trajectories, facilitated by a theoretically-grounded refinement operator that aggregates candidate trajectories from multiple agents. 
    
    \item \textbf{MARINE architecture}. MARINE implements a layered architecture where each refinement layer employs multiple heterogeneous agents operating on a shared reference trajectory. The framework introduces structured trajectory representation, conflict-aware meta-verification, and segment-level integration mechanisms that collectively ensure trajectory improvement without full regeneration. This architecture precisely isolates local improvements while preserving global reasoning coherence. 
    
    \item \textbf{Batch-size optimization}. Theoretical analysis establishes complementary principles for agent orchestration under distinct computational constraints. Under fixed invocation budgets, minimal feasible batches maximize expected performance gains per agent call. Conversely, when no invocation limit exists, logarithmically growing batch schedules guarantee monotonic trajectory improvement with arbitrarily high probability, establishing rigorous worst-case performance bounds critical for reliability-sensitive applications.
    
    \item \textbf{Experimental Validation and Parameter Efficiency}. Comprehensive evaluation on the BrowserComp-ZH benchmark demonstrates that MARINE achieves state-of-the-art performance when implemented with a 685B-parameter LLM. Significantly, an 80B-parameter LLM augmented with MARINE matches the performance of standalone 1000B-parameter models, establishing a new paradigm for parameter-efficient reasoning. The framework consistently surpasses baseline methods including Self-Refine and Best-of-N under equivalent computational constraints, empirically validating the theoretical analysis regarding optimal batch sizing and refinement depth. These results position MARINE as a practical framework for enhancing test-time reasoning capabilities while substantially reducing parameter requirements.
\end{itemize}

\section{Related Work}

\subsection{Multi-Sample Decoding and Test-Time Scaling}

Multi-sample decoding improves LLM reasoning at inference time by drawing multiple candidates and selecting among them. Chain-of-Thought (CoT) prompting exposes intermediate steps and improves performance on arithmetic and symbolic tasks \cite{wei2022chain,zhou2022least,chen2022program}. SC samples many reasoning chains and takes a majority vote over final answers, converting some pass@N potential into higher pass@1 but without formal scaling guarantees \cite{wang2022self}. Subsequent work generalizes multi-sample decoding into BoN selection with reward models or self-certainty signals and shows that inference-time selection can rival or complement post-training alignment \cite{kang2025scalable,park2025ensembling,sun2024fast}. Other work studies SC and BoN under calibrated confidence or budget-aware policies and proposes strategies such as adaptive self-calibration and thought pruning to reduce multi-sample costs \cite{zeng2025revisiting,huang2025efficient,hong2025slim}. \par

These methods treat each sampled trajectory as an independent solution and do not maintain a persistent reference trajectory or analyze monotone trajectory improvement. MARINE instead focuses on a single evolving trajectory and provides a probabilistic analysis of its quality under repeated multi-agent refinement and different batch-size schedules.

\subsection{Self-Refinement and Revision Methods}

Beyond pure sampling, self-refinement methods iteratively critique and revise model outputs. Self-Refine establishes a feedback–refine loop and reports gains without parameter updates \cite{madaan2023self}. Multi-Aspect Feedback (MAF) uses multiple specialized feedback modules targeting distinct error types and improves over Self-Refine on several benchmarks \cite{nathani2023maf}. Structured Reasoning with Revisions (SCREWS) formulates reasoning with revisions via three stages, sampling, conditional resampling and selection, that subsume a range of reflection and tool-use strategies \cite{shridhar2023screws}. \par

Revision is also integrated with preference optimization. Process- and preference-based supervision is combined with search over partial solutions, where reflective-guided exploration yields higher-quality preference data for policy updates \cite{rafailov2023direct,lightman2023let,hao2023reasoning}. Surveys consolidate diverse self-correction strategies and empirical patterns \cite{pan2023automatically}. \par

These approaches are algorithmic and empirical: they typically lack an explicit trajectory metric, do not model how improvement probability decays as solutions strengthen, and do not study optimal batch size or depth under fixed inference budgets. MARINE defines a trajectory-level error, imposes assumptions on local superiority and comparative evaluation, and derives monotone improvement and batch-size trade-offs for structured in-context refinement.

\subsection{Multi-Agent Reasoning}

Multi-agent frameworks deploy multiple instances that collaborate or compete during problem solving. Recent surveys report strong empirical performance of LLM-based multi-agent systems across diverse tasks, highlight their promising prospects, and discuss remaining coordination and reliability challenges \cite{guo2024large,li2024survey,chen2024survey,tran2025multi}.\par

Debate-style systems instantiate multi-agent interaction for reasoning. Prior work shows that single-agent reflection can suffer degeneration of thought and introduces Multi-Agent Debate (MAD) to encourage divergent arguments overseen by a judge \cite{liang2024encouraging}. Follow-up work applies debate to fake-news detection, requirements engineering and stance detection, and uses debate traces as supervision for fine-tuning \cite{subramaniam2025multiagent,li2024can,chan2023chateval}. \par

Existing multi-agent systems generally operate over free-form dialogue, without an explicit shared reference trajectory or trajectory-level guarantees. MARINE instead organizes agents around a common reference trajectory, fuses only locally superior segments under verification, and analyzes how the probability of global improvement scales with the number of exploration agents and the refinement depth.

\begin{figure*}[tb]
\centering
\includegraphics[scale=0.78]{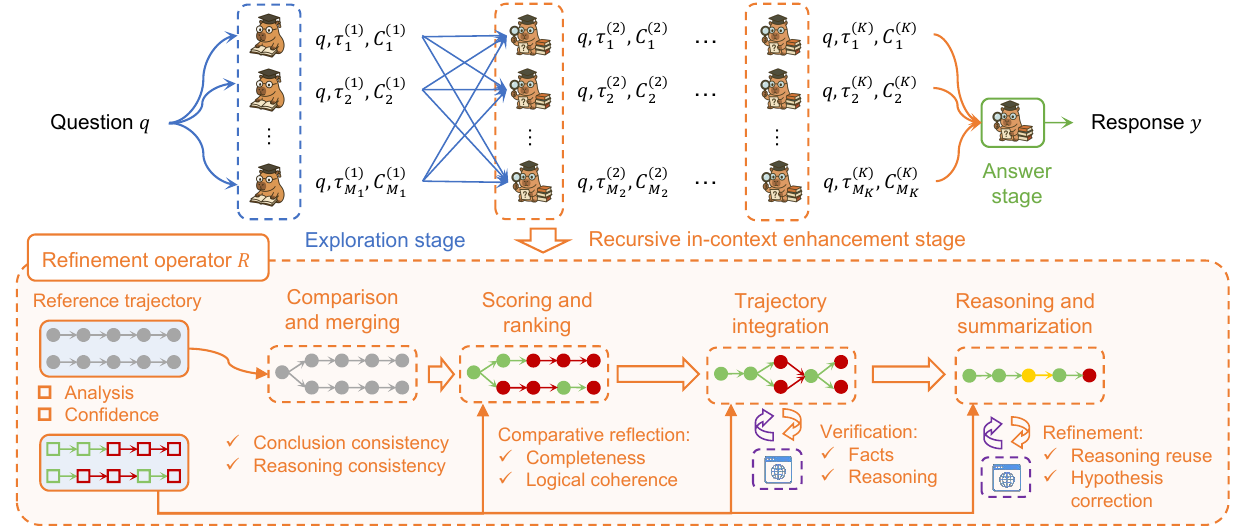}
\caption{\textbf{MARINE Framework: Multi-Agent Recursive IN-Context Enhancement for Trajectory Refinement.} (Top) Layered architecture comprising initial exploration with parallel agents, $K$ recursive enhancement layers propagating reference trajectories as persistent states, and final answer generation. Each layer employs $M_k$ agents operating on structured context $(q,\tau^{(k)},C^{(k)})$ with controlled diversity mechanisms. (Bottom) Refinement operator $R$ workflow: structured trajectory representation, multi-dimensional conflict detection (factual and logical), meta-verification through authority assessment and cross-validation, and segment-level integration of verified improvements. The operator ensures monotonic trajectory improvement via dimensional error minimization while preserving reasoning coherence through hypothesis correction and comparative reflection.}
\label{fig:system}
\vspace{-10pt}
\end{figure*} 

\section{MARINE System}

This section formalizes the trajectory refinement paradigm and delineates the MARINE framework. Unlike one-shot decoding or independent multi-sample selection approaches, MARINE reconceptualizes test-time reasoning as an iterative optimization process centered around a persistent reference trajectory. The framework establishes a theoretically grounded refinement operator that aggregates candidate trajectories from multiple heterogeneous exploration agents. Implementation of this operator employs a trajectory representation methodology and a recursive in-context enhancement mechanism, collectively transforming a base model's pass@N capabilities into pass@1 performance.

\subsection{Trajectory Refinement Paradigm}

Let the original task be $ q \in Q $, and let $ \tau^\star $ denote the ideal reasoning trajectory corresponding to the optimal solution. The standard LLM reasoning process can be abstracted as a fixed-parameter policy $ \pi_\theta $ that generates a complete trajectory $ \tau $ for task $ q $ as
\begin{equation}
\tau \sim \pi_\theta(\cdot \mid q), \quad y = \text{Decode}(\tau), \label{eq:origin_task}
\end{equation}
where $\text{Decode}(\cdot)$ maps the trajectory to the final output $y$. In this formulation, the model attempts to solve the entire task in a single step by sampling directly in a high-dimensional search space. \par

MARINE utilizes an agent-based trajectory refinement operator. Denote the reference trajectories at the $ k $-th round by $ \tau^{(k)} $, and the set of trajectories generated in parallel by $ M_{k+1} $ agents by
\begin{equation}
\tau_i^{(k+1)} \sim \pi_{i}(\cdot \mid q, \tau^{(k)}, C^{(k)}), \quad i \in \left(1,\cdots,M_{k+1}\right)
\end{equation}
where $ C^{(k)} $ encodes supplementary analyses of $ \tau^{(k)} $, such as confidence scores and tool-usage logs. Heterogeneous agents, instantiated via different prompts or reasoning preferences, induce diverse and complementary reasoning paths. The trajectory refinement operator is defined as
\begin{equation}
R\left(q, \tau^{(k)}, C^{(k)}\right) = \left(\tau^{(k+1)},C^{(k+1)}\right). \label{eq:refine_op}
\end{equation}

Under this paradigm, the objective of each reasoning round shifts from ``generating a fully correct trajectory from scratch” to ``performing local refinements around the current reference trajectory.” \par

Trajectory quality is characterized through a continuous distance metric. Each reasoning trajectory $\tau$ is mapped to a $J$-dimensional evaluation vector
\begin{equation}
d(\tau, \tau^\star) = (d_1(\tau, \tau^\star), \dots, d_J(\tau, \tau^\star)) \in [0, 1]^J,
\end{equation}
where each dimension $j$ corresponds to semantic dimensions such as ``whether key facts are correct", ``whether intermediate equations hold", ``whether the logical chain is coherent", etc. The overall distance is defined as:
\begin{equation}
\text{dist}(\tau, \tau^\star) = \frac{1}{J} \sum_{j=1}^J d_j(\tau, \tau^\star) \in [0, 1], 
\end{equation}
where \text{dist} = 0 indicates exact agreement with the ideal trajectory across all dimensions, and \text{dist} = 1 indicates maximal discrepancy with respect to the ideal trajectory.  \par

In an idealized case, progressively improved reference trajectories monotonically increase the distance score on the primary task and yield higher-quality reasoning traces. At the macro level, the refinement problem becomes increasingly localized as $ \tau^{(k)} $ approaches the ideal trajectory $ \tau^\star $. At the micro level, the few-shot context becomes more closely aligned with the ground-truth trajectory, which steers the LLM’s autoregressive generation toward the correct solution and consequently improves accuracy.

\begin{algorithm}[t]
\caption{MARINE Algorithm}
\label{alg:marine}
\begin{algorithmic}[1]
\REQUIRE Query $q$; depth $K$; batch sizes $\{M_k\}_{k=1}^K$
\ENSURE Final answer $y$

\STATE \textbf{Exploration stage}
\STATE Generate initial trajectories $\{(\tau^{(1)}_i, C^{(1)}_i)\}_{i=1}^{M_1}$ in parallel from $M_1$ agents given $q$
\STATE Select reference trajectory $(\tau^{(1)}, C^{(1)})$ from the initial set

\STATE \textbf{Recursive enhancement stage}
\FOR{$k = 1$ to $K$}
    \STATE Generate $\{(\hat{\tau}^{(k)}_i, \hat{C}^{(k)}_i)\}_{i=1}^{M_k}$ in parallel conditioned on $(q, \tau^{(k)}, C^{(k)})$
    \STATE $(\tau^{(k+1)}, C^{(k+1)}) \leftarrow \mathcal{R}(q, \tau^{(k)}, C^{(k)})$
\ENDFOR

\STATE \textbf{Answer stage}
\STATE $y \leftarrow$ generation from an agent based on $(q, \tau^{(K)}, C^{(K)})$
\STATE \textbf{return} $y$
\end{algorithmic}
\end{algorithm}

\subsection{Workflow of MARINE}

As illustrated in Figure \ref{fig:system}, \textbf{the MARINE framework can be analogized to a feedforward neural network of sub-agents}: each layer comprises $M_k$ parallel sub-agents, with reference trajectories $\tau^{(k)}$ propagating ``hidden states" between layers.  \par

At the exploration stage ($k=1$), $M_1$ agents independently and in parallel generate $\{(\tau^{(1)}_1, C^{(1)}_1),\cdots,(\tau^{(1)}_{M_1}, C^{(1)}_{M_1})\}$ based on the initial question $q$. The process then proceeds to the recursive in-context enhancement stage. At the $k$-th round, $M_k$ agents independently execute the function $R$ in \eqref{eq:refine_op} in parallel, producing new $\tau^{(k)}$ and $C^{(k)}$. This stage is recursively executed until the preset limit $k=K$ is reached. In the answer phase, one agent generates the ultimate response to the initial question $q$ based on $\tau^{(K)}$ and $C^{(K)}$. The workflow of MARINE is summarized in Algorithm \ref{alg:marine}. The generation of diverse trajectories in the initial exploration phase and the recursive enhancement of the context are detailed in the following sections.

\subsection{Diverse Trajectory Generation}

At the $k$-th layer, given the tuple $(q,\tau^{(k)},C^{(k)})$, the diversity of the generated trajectories is encouraged through the following mechanisms:

\begin{itemize}[itemsep=0pt, topsep=0pt, parsep=0pt]
    \item \textbf{Reasoning Path Diversity}: Distinct problem-solving preference prompts are applied to different agents. 
    \item \textbf{Intermediate Fact Diversity}: Agents are permitted to invoke different external tools or knowledge sources. 
    \item \textbf{Sampling Strategy Diversity}: By configuring different sampling temperatures, we establish a behavioral spectrum where certain agents favor reliable outputs while others encourage diverse explorations.
\end{itemize}

This design yields generated trajectories $\tau^{(k+1)}$ exhibiting strong local optimality diversity: even when no single trajectory surpasses the reference $\tau^{(k)}$ across all dimensions, superior local-optimum fragments frequently emerge along different evaluation dimensions.

\subsection{Recursive In-Context Enhancement}

To implement the trajectory refinement operator $R$, MARINE enhances the conflict-aware meta-verification mechanism \cite{lu2025co}, which can identify and integrate valuable information from diverse trajectories. As illustrated in Figure \ref{fig:system}, the mechanism comprises four stages:

\begin{enumerate}[itemsep=0pt, topsep=0pt, parsep=0pt]
    \item \textbf{Structured trajectory representation.} All trajectories are first converted into a unified graph-structured representation. This exposes intermediate reasoning steps as aligned nodes, enabling direct comparison across trajectories and avoiding spurious differences.
    \item \textbf{Conflict detection.} The meta-verification module performs a comparative analysis over these structured graphs to merge consistent nodes and detect two types of conflicts: factual conflicts and logical conflicts. 
    \item \textbf{Conflict resolution.} Conflicting results are evaluated based on reliability and rationality: factual nodes are ranked by source authority and external verifiability, while reasoning nodes undergo stress testing via backward substitution and boundary-condition analysis. For unresolved conflicts, the system engages with external verification sources, thereby minimizing the propagation of hallucinated or unsupported content.
    \item \textbf{Trajectory update.} Verified information is recorded in a facts module and used to update the reference trajectory. The update operates at the segment level, replacing erroneous spans with verified facts, inserting missing but necessary intermediate steps, and repairing logical gaps, so that improvements accumulate monotonically across iterations without requiring full regeneration of the trajectory.
\end{enumerate}

\section{Theoretical Analysis of Batch Size} \label{sec:batch}

This section investigates the optimal selection of batch size $M_k$ under different computational constraints, subject to Assumption 4.1.

\begin{assumption}\label{assump:2}
\textbf{Effectiveness of comparative evaluation.} When the number of agents in $k$-th round satisfies $M_k \ge 2$, the evaluation module can identify dimension-wise superiority sources from the candidate set. Specifically, for each dimension $j \in {1, \dots, J}$, it can select from ${\tau^{(k)}} \cup C^{(k)}$ a trajectory segment minimizing the local error $d_j(\cdot,\tau^\star)$.
\end{assumption}

This assumption represents a minimal capability requirement: an evaluation mechanism unable to discern relative trajectory quality would fundamentally lack the capacity to solve the original problem $q$.

\begin{figure*}[tb]
\centering
\includegraphics[scale=0.41]{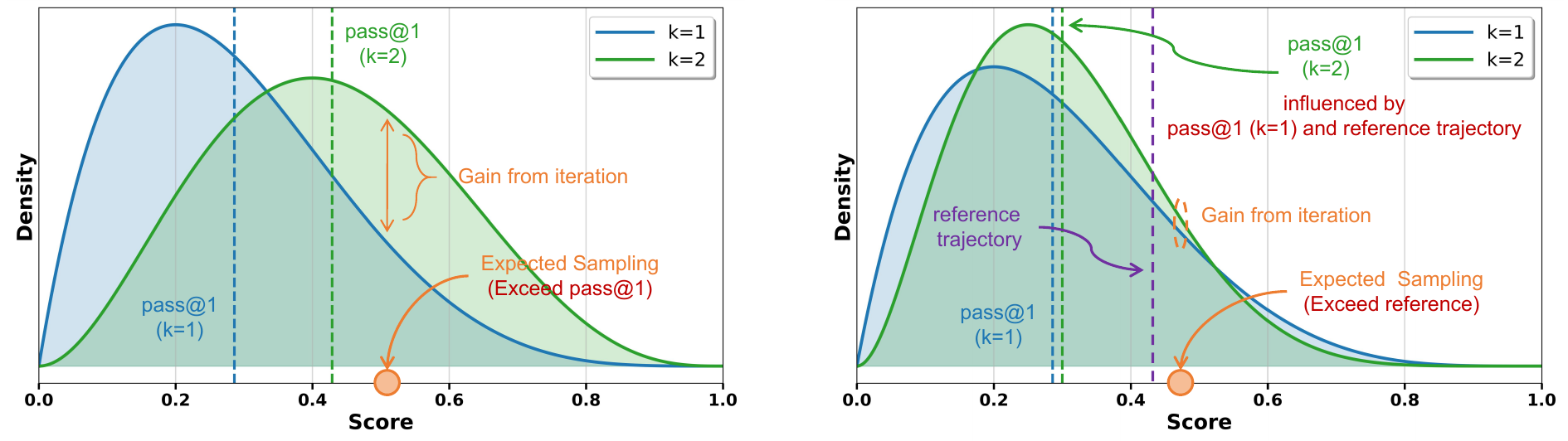} 
\caption{\textbf{Evolution of score distributions for generated solutions under RL (left) and MARINE trajectory refinement (right) across iterations}. The left panel illustrates how parameter updates in RL shift the entire score distribution to higher values, keeping the probability of exceeding the current pass@1 roughly stable. The right panel shows how conditioning on a fixed, increasingly strong reference trajectory in MARINE biases generation toward a high-score region, thereby enlarging the gap between pass@1 and the reference and reducing the probability that newly sampled trajectories surpass the reference over iterations.}
\label{fig:probability}
\vspace{-12pt}
\end{figure*}

\subsection{Comparison with Reinforcement Learning}

A comparative analysis between reinforcement learning (RL) and MARINE elucidates the corresponding decrease in the successful enhancement probability $ p_k $ with increasing iteration count $ k $.  \par

Both MARINE and RL approaches (including GRPO and GSPO variants) leverage groupwise relative comparisons and multi-round iterative procedures. The fundamental distinction resides in their update mechanisms: RL methods perform direct optimization of model parameters, while MARINE operates exclusively within trajectory space through iterative refinement of a reference trajectory via candidate generation and correction.  \par

Figure 2 provides an intuitive illustration of performance evolution contrasting RL (left panel) and MARINE (right panel). The solution score is modeled as a random variable, with its expectation corresponding to the pass@1 metric. In both panels, blue and green distributions represent score distributions at iterations $k=1$ and $k=2$, respectively, with corresponding dashed lines indicating their means (i.e., pass@1 scores). The purple dashed line denotes the score of the reference trajectory selected at iteration $k=1$, assumed to exceed the initial pass@1 score.

In the RL paradigm (left panel), parameter updates shift the entire score distribution rightward as model capabilities improve, thereby increasing the pass@1 score from the blue to green distribution. Under RL frameworks, the probability that a newly sampled solution exceeds the current pass@1 score remains approximately 50\%, as the pass@1 corresponds to the distribution mean. Conversely, the MARINE framework (right panel) demonstrates the effect of conditioning on a fixed reference trajectory. This conditioning induces a bias toward higher-scoring regions, positioning the pass@1 score at $k=2$ (green) between the reference score (purple) and the original pass@1 at $k=1$ (blue). The reference trajectory encodes only partial task information, thereby biasing generation along specific dimensions while model parameters maintain their inherent bias in others. Consequently, the probability of sampling a trajectory that strictly improves upon the reference falls below 50\%. As refinement progresses and the reference trajectory strengthens, the gap between its score and the corresponding pass@1 widens, further reducing the probability of sampling strictly superior trajectories. This dynamic results in diminishing improvement rates across iterations for MARINE, whereas RL maintains a consistently high improvement probability.

Formally, at the $k$-th iteration, given the reference trajectory $\tau^{(k)}$ with score $r_k := S(\tau^{(k)})$, consider $M_k$ generated trajectories $\{\tau_i^{(k+1)}\}_{i=1}^{M_{k+1}}$ with scores $\{S_i^{(k+1)}\}_{i=1}^{M_{k+1}}$. Define the probability that a single candidate surpasses the reference as
\begin{equation}
    p_k := \mathbb{P}\big(S^{(k+1)}_i > r_k\big), \quad i = 1, \ldots, M_{k+1} . \label{eq:impro_prob}
\end{equation}

It leads to the following theoretical characterization:

\begin{proposition}[Progressively decreasing probability of successful sampling in MARINE] \label{pro:1}
In MARINE-based trajectory optimization, iterative refinement of the reference trajectory increases the scores of generated trajectories, but the gap between the pass@1 score and the reference trajectory score grows monotonically with the number of iterations, which in turn causes the probability that a generated trajectory attains a higher score than the reference to decrease steadily.
\end{proposition}

\subsection{Optimal Batch Size with Fixed Budget}

This section considers the setting where the total number of agent invocations $T$ is bounded. At the $k$-th round, the focus is whether at least one success occurs among the $M_k$ parallel invocations. In this case, Theorem \ref{theo:2} holds.

\begin{theorem}[Optimal exploring batch size $M_k$ under constrained agent invocation budgets $T$]  \label{theo:2}
Suppose the total number of agent invocations $T$ is fixed and $p_k \in (0, 0.5)$ for all rounds $k$. Under this total invocation constraint, a smaller batch size $M$ yields higher efficiency in terms of expected gain per invocation. In particular, if the MARINE framework requires $M_k \geq 2$ to ensure sufficient complementarity among agents and satisfy Assumption \ref{assump:2}, the optimal choice is
\begin{equation} 
M^\star_k = M_{\min} = 2.
\end{equation}
\end{theorem}
\begin{proof}
    The proof is provided in Appendix \ref{app:proof-theo2}.
\end{proof} \vspace{-15pt}

\subsection{Batch Size with Unlimited Budget}

The previous subsection analyzed the optimal batch size under a fixed $T$. This subsection considers a setting without the invocation limit and studies how small the per-iteration batch size can be while still guaranteeing monotone improvement of the reference trajectory with high probability. \par

The core question is whether there exists a schedule $\{M_k\}_{k=1}^{\infty}$ such that, from some iteration onward, the reference trajectory improves at every step with probability arbitrarily close to one. The following theorem shows that a logarithmically increasing batch size is sufficient, provided the per-sample improvement probability does not vanish. \par

\begin{theorem}[Monotone improvement with growing batch size] \label{theo:3}
Suppose there exists a constant $\underline{p} > 0$ and an index $\tilde{k}$ such that $p_k \ge \underline{p}$ for all $k \ge \tilde{k}$. If the batch sizes satisfy
\begin{equation}
M_k \ge \frac{-2 \ln k}{\lvert \ln(1 - \underline{p}) \rvert},
\quad k \ge \tilde{k},
\end{equation}
then for any $\delta > 0$ there exists $\tilde{k}' \ge \tilde{k}$ such that, with probability at least $1 - \delta$, every iteration $k \ge \tilde{k}'$ produces at least one trajectory that strictly improves over the current reference trajectory. In other words, under a mild lower bound on the per-sample improvement probability, a slowly increasing batch size of order $\mathcal{O}(\log k)$ suffices to make ``continuous” improvement overwhelmingly likely in the limit.
\end{theorem}  
\begin{proof}
    The proof is provided in Appendix \ref{app:proof-theo3}.
\end{proof}  \vspace{-12pt}

\textbf{Theorems~\ref{theo:2} and~\ref{theo:3} establish complementary theoretical foundations for inference-time resource allocation.} Theorem~\ref{theo:2} provides the optimal batch-size configuration under fixed invocation budgets, maximizing expected performance gains through minimal feasible batches. This principle is essential for practical deployments where computational resources are strictly constrained yet average-case performance remains the primary objective. Conversely, Theorem~\ref{theo:3} establishes rigorous worst-case guarantees by prescribing logarithmically growing batch schedules that bound performance degradation probability below $\delta$ while minimizing computational overhead. This theoretical guarantee becomes critical in reliability-sensitive applications where consistent monotonic improvement outweighs average performance considerations. Together, these theorems delineate a complete optimization framework for test-time reasoning systems across diverse operational constraints.

\section{Experimental Evaluation and Ablations}

\subsection{Experimental Setup}

In this section, the BrowserComp-ZH benchmark \cite{zhou2025browsecomp} is used as a representative multi-hop retrieval task. It is a Chinese-focused evaluation dataset that measures model performance in complex information-seeking scenarios, where each query requires several reasoning steps over a large document collection to gather supporting evidence and synthesize an answer, closely mirroring the information search process in a browser environment.


To evaluate the generality of the proposed method, experiments are conducted on two classes of language models spanning large and small scales. The large-scale model is DeepSeek-V3.2 (without thinking), with a total parameter count of 685B and 37B active parameters. The small-scale model is Qwen3-next-80B-A3B in thinking mode, with 30B total parameters and 3B active parameters. All models are open-source, which ensures strict reproducibility of the experimental results. Besides, all reported results are obtained by averaging over three independent samples to mitigate stochastic fluctuations. \par

\begin{figure}[tb]
\centering
\includegraphics[scale=0.19]{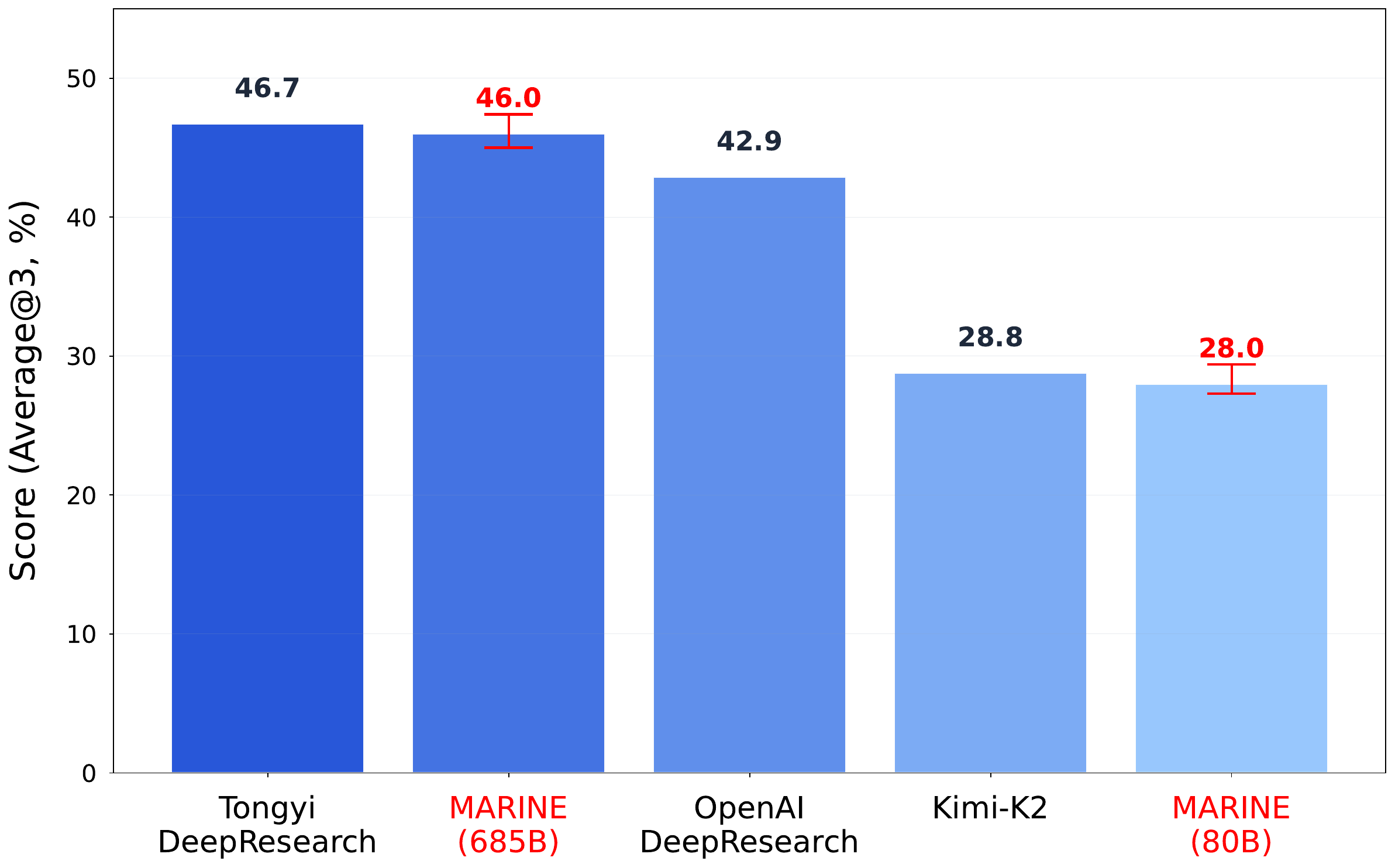}
\caption{\textbf{Performance Comparison.} MARINE achieves SOTA with the 685B LLM and matches Kimi-K2 with the 80B LLM.}
\label{fig:performance}
\vspace{-12pt}
\end{figure}

\subsection{Main Results on BrowserComp-ZH}

Following Theorem \ref{theo:2}, MARINE is instantiated with four refinement rounds $K = 4$ and a constant batch size $M_k = 2$, which maximizes the expected gain per agent invocation under a fixed budget while preserving sufficient intra-batch diversity for reliable comparative evaluation.  \par

As shown in Figure \ref{fig:performance}, MARINE with the 685B LLM achieves a 46.0\% pass@1 score on the BrowserComp-ZH task with a matched invocation budget, outperforming strong baselines. This improvement arises from the strategic allocation of the budget to a structured multi-agent refinement process, centered around a persistent reference trajectory. The iterative refinement progressively reduces trajectory-level errors by targeting contested regions, excelling in multi-hop retrieval, and mitigating issues where intermediate evidence errors constrain single-shot decoding. While MARINE’s result of 46.0\% is slightly below Tongyi DeepResearch’s 46.7\%, it offers a generalizable framework that avoids the training overhead inherent in Tongyi’s agentic RL process. Notably, MARINE achieves a performance with the 80B-A3B model, matching Kimi-K2’s results, despite the significantly smaller model size.

\subsection{Performance Comparison with Baseline Methods}

In this section, MARINE's performance is evaluated against baseline methods, specifically single-sample CoT, Self-Refine, and BoN under a fixed agent invocation budget. The results highlight MARINE’s superior ability to leverage iterative multi-agent trajectory refinement. \par

Table \ref{tab:agents_method} presents the performance of MARINE across different LLM sizes under a fixed total agent invocation budget of $T + 1 = 9$ (average@3). MARINE achieves the highest pass@1 accuracy compared to baseline methods. Specifically, with the 685B model, MARINE reaches a pass@1 score of 46.0\%, surpassing BoN at 35.3\%, Self-Refine at 40.5\%, and single-sample CoT at 26.0\%. Similarly, with the 80B model, MARINE achieves 28.0\%, outperforming BoN (22.8\%), Self-Refine (11.1\%), and CoT (10.0\%). These results highlight MARINE's effectiveness in leveraging multiple agents for iterative reference trajectory refinement, which enhances the reasoning process by isolating contexts and iteratively improving the quality of candidate trajectories. Compared with Self-Refine, MARINE employs contrastive reflection to select promising partial trajectories within each batch, ensuring more precise trajectory corrections. Furthermore, MARINE integrates complementary information from candidates across the batch, rather than relying solely on intra-batch diversity, thereby consistently outperforming BoN under the same computational budget.

\begin{table}[t]
  \centering
  \caption{Score under a fixed total agent invocation budget $T +1= 9$ (average@3, \%).}
  \label{tab:agents_method}
  \begin{tabular}{ccccc}
    \hline
    Method & MARINE & BoN & Self-Refine & CoT \\
    \hline
    685B & \textbf{46.0}$\uparrow$ & 35.3 & 40.5 & 26.0 \\
    80B & \textbf{28.0}$\uparrow$ & 22.8 & 11.1 & 10.0 \\
    \hline
  \end{tabular}
  \vspace{-12pt}
\end{table}

\subsection{Ablation Study on Batch Size}

Table \ref{tab:agents_batch} presents the pass@1 scores for the MARINE framework under a fixed total agent invocation budget of $ T + 1 = 9 $, with varying batch sizes $ M_k $ ranging from 1 to 8, for both 685B and 80B language models. For the 685B model, the highest pass@1 score, 46.0\%, is achieved with a batch size of $ M_k = 2 $, while smaller (1) and larger (4, 8) batch sizes yield lower scores of 40.5\%, 41.2\%, and 39.4\%, respectively. For the 80B model, the best performance is again obtained with $ M_k = 2 $, yielding a score of 28.0\%, with scores for $ M_k = 1 $ dropping to 11.1\%, and larger batch sizes showing diminishing returns (24.9\% for $ M_k = 4 $, 22.8\% for $ M_k = 8 $). \par

These results reinforce the importance of batch size selection in the MARINE framework. The optimal batch size of $ M_k = 2 $ maximizes the pass@1 performance, aligning with the theoretical prediction that, under a fixed invocation budget, the expected improvement per invocation is maximized by the smallest feasible batch size. Smaller batch sizes are ineffective due to the lack of sufficient diversity in candidate trajectories, which prevents the refinement process from reliably identifying improvements. On the other hand, larger batch sizes (4 and 8) lead to diminishing returns as the marginal gain per agent call decreases. This is consistent with the theoretical trade-off between exploration width and refinement depth, where larger batches increase the candidate pool and extend the context, making it harder to isolate relevant improvements and amplifying the impact of noisy or redundant content. Overall, the results empirically validate the theoretical insights, highlighting that the optimal batch size balances exploration and refinement efficiently, thus maximizing performance within the given budget.

\begin{table}[t]
  \centering
  \caption{Effect of batch size $M_k$ on MARINE pass@1 under a fixed total agent invocation budget $T+1 = 9$ (average@3, \%).}
  \label{tab:agents_batch}
  \begin{tabular}{ccccc}
    \hline
    $M_k$ & $1$ & $2$ & $4$ & $8$ \\
    \hline
    685B & 40.5 & \textbf{46.0}$\uparrow$ & 41.2 & 39.4 \\
    80B & 11.1 & \textbf{28.0}$\uparrow$ & 24.9 & 22.8 \\
    \hline
  \end{tabular}
  \vspace{-12pt}
\end{table}

\subsection{Ablation Study on Refinement Rounds}

Table 3 presents the pass@1 scores for the MARINE framework across varying refinement rounds $ K $ under a fixed total agent invocation budget of $ T = 2K + 1 $. For the 685B model, MARINE shows consistent improvement, reaching 46.0\% at $ K = 4 $, outperforming Self-Refine, which reaches 40.5\%.  The pass@N baseline exceeds MARINE at 56.4\% by $ K = 4 $. For the 80B model, MARINE demonstrates a particularly strong advantage, surpassing pass@N at ( K = 1 ) with 22.1\% compared to 19.0\%. At $ K = 4 $, MARINE achieves 28.0\%, outperforming pass@N's 32.9\%. Self-Refine lags behind, remaining at 11.1\% across all rounds.

These results highlight the superior performance of MARINE, particularly at smaller $ K $ values and for smaller models, where it not only outperforms Self-Refine but also exceeds pass@N. While pass@N eventually outperforms MARINE, the framework excels in efficiency, showing competitive performance in resource-constrained scenarios. These findings emphasize the effectiveness of early refinement rounds in improving pass@1, while also demonstrating the diminishing returns as the number of rounds increases, especially for smaller models where the impact of each additional round is less pronounced.

\begin{table}[t]
  \centering
  \caption{Effect of refinement rounds $K$ under matched total agent invocations $2K + 1$ for MARINE, Self-Refine and oracle style pass@N (average@3, \%).}
  \label{tab:agents_iteration}
  \begin{tabular}{cccccc}
    \hline
    $K$ & $1$ & $2$ & $3$ & $4$ \\
    \hline
    MARINE (685B)  & 35.6 & 41.2 & 44.3 & 46.0 \\
    Self-Refine (685B) & 32.9 & 37.0 & 39.4 & 40.5 \\
    pass@N (685B) & \textbf{44.3}$\uparrow$ & \textbf{49.5}$\uparrow$ & \textbf{54.0}$\uparrow$ & \textbf{56.4}$\uparrow$ \\ \hline
    
    MARINE (80B)  & \textbf{22.1}$\uparrow$ & \textbf{26.0}$\uparrow$ & 27.0 & 28.0 \\
    Self-Refine (80B) & 11.1 & 10.0 & 11.1 & 11.1 \\
    pass@N (80B) & 19.0 & 26.0 & \textbf{30.8}$\uparrow$ & \textbf{32.9}$\uparrow$ \\
    \hline
  \end{tabular}
  \vspace{-12pt}
\end{table}

\section{Conclusion}

MARINE establishes a framework reconceptualizing test-time reasoning as iterative refinement of persistent reference trajectories via multi-agent collaboration, systematically transforming base models' pass@N capabilities into reliable pass@1 performance.  Theoretical analysis demonstrates minimal feasible batches maximize performance under fixed invocation budgets, while logarithmically growing schedules guarantee continuous improvement.  Empirical validation reveals unprecedented parameter efficiency: an 80B-parameter model augmented with MARINE matches standalone 1000B-parameter systems, reducing parameter requirements by over an order of magnitude.  The recursion-based architecture overcomes single-agent limitations through structured trajectory representation and conflict-aware integration.  This paradigm fundamentally decouples reasoning capability from parameter scale, establishing a new approach to parameter-efficient LLM deployment.  Crucially, under a fixed computational budget, MARINE yields higher-quality samples for alignment and optimization workflows than traditional sampling-and-ranking strategies (i.e., the scenario where $k=1$). Hence, it exhibits significant potential for enhancing post-training efficiency.

\section*{Impact Statement}
MARINE establishes a theoretically-grounded framework for enhancing the reliability and efficiency of LLM-based agents through recursive trajectory refinement rather than single-pass decoding or unstructured multi-sampling.  Rigorous analysis demonstrates that minimal feasible batches maximize expected performance under fixed invocation budgets, while logarithmically growing batch schedules guarantee monotonic trajectory improvement with arbitrarily high probability—establishing critical worst-case performance bounds for reliability-sensitive applications. Empirical validation confirms unprecedented parameter efficiency, with smaller models augmented by MARINE matching or exceeding the performance of substantially larger standalone systems. Notably, this methodology optimizes reasoning trajectories under fixed computational budgets and exhibits substantial potential to enhance post-training efficiency. Specifically, it generates samples of higher quality than those produced by conventional sampling-and-ranking strategies (i.e., $k=1$) for alignment and optimization procedures. Consequently, this capability can reduce both data volume and computational requirements during subsequent fine-tuning phases. While these advances promise more capable and efficient reasoning systems, they also introduce risks, including potential misinformation propagation and strategic manipulation.  Therefore, responsible deployment necessitates robust safeguards such as comprehensive auditing of external tools and evaluation signals, continuous monitoring for failure modes, and explicit constraints on high-stakes applications.

\bibliography{MARINE/refer}
\bibliographystyle{icml2025}

\newpage
\appendix
\onecolumn

\section{Proof of Theorem \ref{theo:2}} \label{app:proof-theo2}

Since the candidates are independently and identically distributed within the same round, the probability of obtaining at least one improvement in the $k$-th round is
\begin{align}
P_{\text{succ}}(M_k, k)
&:= \mathbb{P}(\mathcal{E}_k) \\
&= 1 - \mathbb{P}\left(\forall i,\ S^{(k)}_i \le r_k\right) \\
&= 1 - (1 - p_k)^{M_k},
\end{align}
where $p_k$ is defined in \eqref{eq:impro_prob}. The expected improvement in reference quality at the $k$-th round is
\begin{align}
\mathbb{E}[r_{k+1} - r_k] 
& = g_k \cdot P_{\text{succ}}(M_k, k) \\ 
& = g_k \left( 1 - (1 - p_k)^{M_k} \right),
\end{align}
where $g_k$ denotes the expected gain conditional on success in round $k$.  

To compare the cost effectiveness of different batch sizes $M_k$, define the expected gain per invocation as
\begin{align}
h_k(M_k) 
& := \frac{\mathbb{E}[r_{k+1} - r_k]}{M_k}  \\
& = g_k \cdot \frac{1 - (1 - p_k)^{M_k}}{M_k}.
\end{align}
Since $g_k$ does not depend on $M_k$, it suffices to study
\begin{equation}
\tilde{h}_k(M_k) 
:= \frac{1 - (1 - p_k)^{M_k}}{M_k}.
\end{equation}

The following simple calculus fact will be useful. For any fixed 
$p_k \in (0, 0.5)$, consider
\begin{equation}
\tilde{h}(M_k) := \frac{1 - (1 - p_k)^{M_k}}{M_k},
\end{equation}
where $M_k \in \mathbb{Z}^+$ and $M_k \geq 2$. Increasing $M_k$ raises the probability that at least one agent succeeds in a given round, but the marginal benefit per invocation diminishes, which is exactly captured by the decrease of $\tilde{h}(M_k)$ in $M_k$. 

Extending $M_k$ to a real variable $x > 0$, set $a := 1 - p_k \in (0.5, 1)$ and define $f(x) := 1 - a^x$ for $x > 0$. Then
\begin{equation}
\tilde{h}(x) = \frac{f(x)}{x}, \quad x > 0.
\end{equation}
The first derivative of $\tilde{h}$ is
\begin{equation}
\tilde{h}'(x) = \frac{a^x(1 - x \ln a) - 1}{x^2}.
\end{equation}
The sign of $\tilde{h}'(x)$ is determined by
\begin{equation}
N(x) := a^x(1 - x \ln a) - 1.
\end{equation}
Write $b := - \ln a > 0$ and note that $a^x = e^{-b x}$. Then
\begin{equation}
N(x) = e^{-b x}(1 + b x) - 1.
\end{equation}
Define $g(t) := e^{-t}(1 + t)$ with $t := b x > 0$. Its derivative satisfies
\begin{equation}
g'(t) = - t e^{-t} < 0 \quad \text{for all } t > 0,
\end{equation}
so $g(t)$ is strictly decreasing on $(0, \infty)$. Since $\lim_{t \to 0^+} g(t) = 1$, it follows that $g(t) < 1$ for all $t > 0$, and therefore
\begin{equation}
N(x) = g(t) - 1 < 0 \quad \text{for all } x > 0.
\end{equation}
Consequently $\tilde{h}'(x) < 0$ for all $x > 0$, so $\tilde{h}(x)$ is strictly decreasing on $(0, \infty)$. Under Assumption \ref{assump:2}, it follows that $\tilde{h}(x)$ is strictly decreasing at the integer points $M_k=2,3,\cdots$, which implies

\begin{equation}
\tilde{h}(2) > \tilde{h}(3) > \tilde{h}(4) > \dots
\end{equation}
for integer $M_k \geq 2$, which proves Theorem \ref{theo:2}.

\section{Proof of Theorem \ref{theo:3}} \label{app:proof-theo3}

Let $\mathcal{F}_k$ denote the event that no corrected trajectory surpasses the reference trajectory in the $k$-th iteration, with probability $\mathbb{P}(\mathcal{F}_k)$. The core question is whether, for sufficiently large $k$, the probability that almost every iteration improves the reference trajectory from some round onward approaches 1. To formalize this, consider iterations starting from a threshold $\tilde{k}$:  
\begin{equation}
k = \tilde{k}, \tilde{k}+1, \dots
\end{equation}
We aim for the probability of the event
\begin{equation}
\mathcal{F}_{\tilde{k}} \cup \mathcal{F}_{\tilde{k}+1} \cup \dots
\end{equation}
to be sufficiently small, i.e., the probability that failure occurs beyond round $\tilde{k}$ with negligible likelihood approaches 1. By the union bound (Boole's inequality): 
\begin{equation}
\mathbb{P}\Big( \bigcup_{k=\tilde{k}}^{\infty} \mathcal{F}_k \Big) \le \sum_{k=\tilde{k}}^{\infty} \mathbb{P}(\mathcal{F}_k) = \sum_{k=\tilde{k}}^{\infty} (1 - p_k)^{M_k}.
\label{eq:boole_uneq}
\end{equation}

Thus, if $M_k$ can be chosen such that the tail sum $\sum_{k=\tilde{k}}^{\infty} (1 - p_k)^{M_k}$ is sufficiently small, the probability of no failure after round $K$ converges to 1. Assume that \( p_k \ge \underline{p} > 0 \) for all \( k \) (or at least after a certain number of rounds), which implies:

\begin{equation}
(1 - p_k)^{M_k} \le (1 - \underline{p})^{M_k}.
\end{equation}

Then, select a target sequence \( \varepsilon_k \) with a sufficiently fast decay rate, e.g.,

\begin{equation}
\varepsilon_k := \frac{1}{k^2}, 
\end{equation}
and require that for each \( k \), \( M_k \) satisfies:

\begin{equation}
(1 - \underline{p})^{M_k} \le \varepsilon_k = \frac{1}{k^2}. 
\end{equation}

This is equivalent to:

\begin{equation}
M_k \ge \frac{\ln \varepsilon_k}{\ln(1 - \underline{p})} = \frac{-2 \ln k}{|\ln(1 - \underline{p})|}. 
\end{equation}

Therefore, \( M_k \) must grow at least at the rate of \( \mathcal{O}(\log k) \). With this choice, we have:

\begin{equation}
\sum_{k=1}^{\infty} (1 - p_k)^{M_k} \le \sum_{k=1}^{\infty} (1 - \underline{p})^{M_k} \le \sum_{k=1}^{\infty} \frac{1}{k^2} < \infty. 
\end{equation}

Hence, based on equation \eqref{eq:boole_uneq}, we conclude that for any \( \tilde{k} \), the probability of failure occurring from round \( \tilde{k} \) onward has a controllable upper bound:
\begin{equation}
\mathbb{P}\Big( \bigcup_{k=\tilde{k}}^{\infty} \mathcal{F}_k \Big) \le \sum_{k=\tilde{k}}^{\infty} \frac{1}{k^2} \to \frac{\pi^2}{6} - \sum_{k=1}^{\tilde{k}-1} \frac{1}{k^2} \quad \text{(series sum of the Basel problem)}.
\end{equation}

Next, define \( \mathcal{G}_{\tilde{k}'} = \bigcap_{k=\tilde{k}'}^{\infty} \overline{\mathcal{F}}_k \) (where \( \overline{\mathcal{F}}_k \) denotes the non-failure event at round \( k \)), so that 
\begin{equation}
    \mathbb{P}(\mathcal{G}_{\tilde{k}'}) = 1 - \mathbb{P}\left( \bigcup_{k=\tilde{k}'}^{\infty} \mathcal{F}_k \right).
\end{equation}

Since \( \sum_{k=1}^{\infty} \frac{1}{k^2} < \infty \), by the tail property of convergent series, for any \( \delta > 0 \), there exists \( \tilde{k}' \ge \tilde{k} \) such that \( \sum_{k=\tilde{k}'}^{\infty} \frac{1}{k^2} < \delta \). Combining Boole's inequality with \( \mathbb{P}(\mathcal{F}_k) \le \frac{1}{k^2} \), it obtain:

\begin{equation}
    \mathbb{P}\left( \bigcup_{k=\tilde{k}'}^{\infty} \mathcal{F}_k \right) \le \sum_{k=\tilde{k}'}^{\infty} \frac{1}{k^2} < \delta.
\end{equation}

Thus, $ \mathbb{P}(\mathcal{G}_{\tilde{k}'}) \ge 1 - \delta $. This establishes Theorem \ref{theo:3}.

\end{document}